\documentclass[%
,twocolumn%
,secnumarabic%
,amssymb,aps,prb,nobibnotes]{revtex4-1}
\usepackage{docs}
\usepackage{graphicx}
\usepackage{xcolor}
\expandafter\ifx\csname package@font\endcsname\relax\else
 \expandafter\expandafter
 \expandafter\usepackage
 \expandafter\expandafter
 \expandafter{\csname package@font\endcsname}%
\fi
\DeclareRobustCommand\substyle{\name@idx{document substyle}}%
\DeclareRobustCommand\classoption{\name@idx{document class option}}%
\DeclareRobustCommand\classname{\name@idx{document class}}%
\def\name@idx#1#2{%
 {\ttfamily#2}%
 \index{#2\space#1=\string\ttt{#2}\space#1}\index{#1>#2=\string\ttt{#2}}%
}%

\begin{document}
\title{Single-edge transport in an InAs/GaSb quantum spin Hall insulator}%

\author{Fran\c cois Cou\"{e}do, Hiroshi Irie, Kyoichi Suzuki, Koji Onomitsu, and Koji Muraki}
\email{muraki.koji@lab.ntt.co.jp}
\affiliation{NTT Basic Research Laboratories, NTT Corporation, 3-1 Morinosato-Wakamiya, Atsugi, Kanagawa 243-0198, Japan}

\begin{abstract}

We report transport measurements in a single edge channel of an InAs/GaSb quantum spin Hall insulator, where the conduction occurs through only one pair of counterpropagating edge modes. By using a specific sample design involving highly asymmetric current
paths, we electrically isolate a single edge channel of the two-dimensional topological insulator from the other edge. This enables us to probe a single edge by multiterminal measurements. Both two-terminal and four-terminal resistances show a nearly quantized plateau around $h/e^2$ for a 4-$\mu$m-long edge, indicating quasiballistic transport. Our approach is advantageous in that it allows us to gain insight into a microscopic region from local measurements. 

\end{abstract}
\maketitle


Boundaries of topological insulators (TIs) provide a new platform to study low-dimensional electron systems with unique properties derived from the non-trivial bulk band structure of the parent TI\cite{hasan2010,qi2011}. Two-dimensional (2D) TIs, also known as quantum spin Hall insulators (QSHIs) \cite{kane2005,bernevig2006,liu2008}, support on their edges a one-dimensional (1D) electron system, often referred to as a \textquotedblleft helical liquid\textquotedblright\ \cite{wu2006}, which possesses peculiarities not shared by conventional 1D conductors such as carbon nanotubes and semiconductor nanowires. A salient feature of a helical liquid is that it comprises only one Kramers pair; that is, there are only one left-moving and one right-moving modes, which are time-reversal conjugates of each other, with reversed spin and momentum directions. Under zero magnetic field, backscattering by nonmagnetic impurities is forbidden by time-reversal symmetry, from which ballistic transport with quantized conductance is expected. Coulomb interaction drives such a system into an even more exotic state described as a helical Luttinger liquid \cite{xu2006, li2015,chou2015}. The reduced degree of freedom and spin-momentum locking characteristic of helical edge states have also led to various proposals \cite{fu2009,beenakker2013,mi2013,crepin2014,lee2014} and experiments \cite{knez2012,hart2014,pribiag2015,shi2015,bocquillon2016} to explore exotic phenomena such as topological superconductivity.

Transport studies of helical liquids based on QSHIs involve two aspects
specific to this system that need to be taken into account: the presence of the
2D bulk region and the edge channels on opposite sides. Previous studies
have established appropriately designed HgTe/CdTe \cite{konig2007,roth2009,brune2012,olshanetsky2015} and InAs/GaSb \cite{knez2011,suzuki2013,knez2014,du2015,suzuki2015,qu2015,mueller2015} quantum wells as 2D TIs, and ballistic transport has been reported for sufficiently short
edges \cite{konig2007,roth2009,du2015,olshanetsky2015}. Beyond this mesoscopic regime, however, edge resistance is observed to increase with edge length \cite{knez2014,du2015}, indicating the existence of a process that
equilibrates the counterpropagating modes. Even in mesoscopic samples
exhibiting a conductance plateau close to the expected quantized value,
deviation from perfect quantization is discernible, and the conductance usually shows
fluctuations as the Fermi level is swept across the bulk energy gap. Various scenarios have been proposed to explain the dissipation in long helical edge
channels \cite{maciejko2009,strom2010,schmidt2012,lunde2012, maestro2013,vayrynen2013,altshuler2013} and the conductance fluctuations in the mesoscopic regime \cite{cheianov2013,vayrynen2013}, where
coupling with the 2D region is often conjectured as a key ingredient \cite{vayrynen2013,zhang2014}. Since
electrical current entering the source (drain) contact is split into two
paths along the opposite edges, scattering at one edge affects the current
partition and hence the voltage on the other edge. Thus, it is highly
desirable to perform transport measurements in a simpler setup where the source
and drain contacts are connected by a single conducting path, as in the case
of conventional 1D conductors. 

In this paper, we demonstrate the realization of effective
single-edge transport measurements on an InAs/GaSb QSHI. We employ a device geometry in which the current flows mostly along one edge in the bulk insulating regime as a result of highly asymmetric
current path lengths. The device geometry is also useful for distinguishing the contribution of
bulk and edge transport. Using a dual-gate configuration, we demonstrate a
crossover from bulk-dominant to single-edge transport regimes tuned by an
electric field. In the mesoscopic regime with a short edge length of $4$ $\mu
$m, we observe a conductance plateau near the value expected for a single
helical edge channel. Comparison between two-terminal and four-terminal
measurements provides insight into the roles of the contacts. 

\begin{figure}[!t]
 \centering
 \includegraphics[scale=0.65]{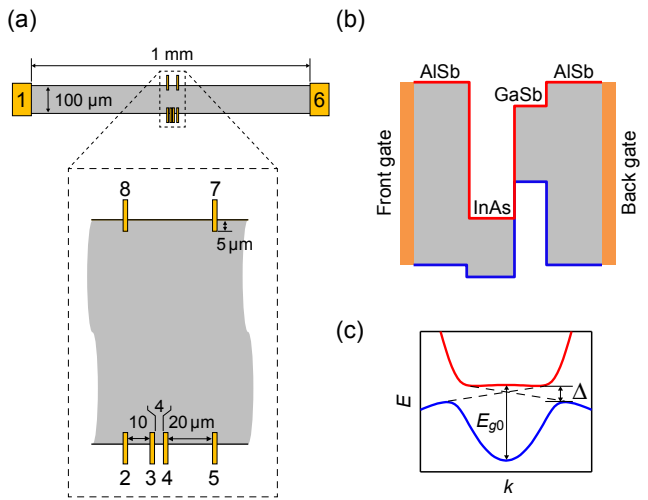}
   \caption{(a) Schematic representation of the sample design. The close-up represents the main part of the device probed by transport measurements. (b) Schematics of band-edge profile of the heterostructure along the growth direction. (c) Illustration of the energy band diagram for inverted InAs/GaSb heterostructures. $\Delta$ is the hybridization gap and $E_{g0}$ is the degree of band overlap. \label{fig1}}
 \end{figure}

   Figure \ref{fig1}(a) schematically shows the device geometry we used for
single-edge transport measurements. The active region of the device with
an InAs/GaSb heterostructure has a $1$-mm-long and $100$-$\mu$m-wide rectangular mesa shape. In addition to the large Ohmic contacts ($1$ and $6$) at
the ends of the mesa, small $2$-$\mu$m-wide contacts are attached to the lower
and upper edges near the center of the mesa, at different spacings ($4$, $10$,
and $20$ $\mu$m) as shown in the enlarged view. We perform single-edge
transport measurements by passing current between two contacts on the lower
edge, for example, between $2$ and $5$. In the QSHI phase, the current flowing
from $2$ to $5$ can in principle take two paths along the mesa edge. However,
the total edge length of $\sim2$ mm for the longer path, $2$-$1$-$8$-$7$-$6$-$5$, is nearly two orders of magnitude greater than that for the
shorter path, $2$-$3$-$4$-$5$. Consequently, we expect the
conduction along the longer edge to be negligible in the QSHI phase.

The undoped InAs/GaSb heterostructure we studied comprises InAs (top) and GaSb
(bottom) layers with nominal thicknesses of 12 and 8 nm, respectively,
sandwiched between 50-nm-thick AlSb barriers. For this InAs layer thickness, the system is
in the band-inverted regime, where the bottom of the InAs electron subband is
located below the top of the GaSb hole subband [Fig.~\ref{fig1}(b)] \cite{altarelli1983}. Hybridization of electron and
hole wave functions through the heterointerface opens a hybridization gap $\Delta$ in
the bulk energy spectrum [Fig.~\ref{fig1}(c)] \cite{altarelli1983,yang1997,lakrimi1997}. We use a dual-gate configuration to independently tune the relative alignment of the electron and hole subbands and the position of the Fermi level $E_{\textrm{\scriptsize{F}}}$. As we will show below, this dual-gate configuration is essential in order to suppress the residual bulk conduction and thereby realize single-edge transport. The sample was processed by photolithography and wet etching. Ti/Au Ohmic contacts were evaporated after the GaSb cap and upper AlSb barrier had been selectively etched
down to the InAs layer. The Ti/Au front gate, which covers the entire mesa
including the contact regions, was evaporated after atomic layer deposition of
a 25-nm-thick Al$_{2}$O$_{3}$ gate dielectric. The $n^{+}$-GaAs substrate is
used as a back gate. Transport measurements are performed using lock-in
techniques, with a sufficiently low current ($\leq1$ nA), at temperature
$T=0.25$ K unless otherwise stated. In the following, $R_{ij,kl}$ indicates
the resistance obtained by driving current from contact $i$ to $j$ and
measuring the resulting voltage $V_{kl}$ between probes $k$ and $l$.


  \begin{figure}
  \includegraphics[scale=0.55]{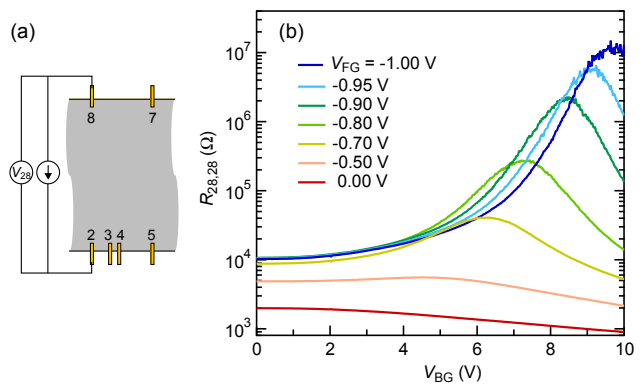}
  \caption{(a) Schematic representation of the measurement configuration for the two-terminal resistance $R_{28,28}$ used to probe the bulk conduction. (b) Back-gate voltage ($V_{\textrm{\scriptsize{BG}}}$) dependence of $R_{28,28}$ at different fixed front-gate voltage  $V_{\textrm{\scriptsize{FG}}}$.\label{fig2}}
  \end{figure}   
  
    \begin{figure*}
 \centering
  \includegraphics[scale=0.58]{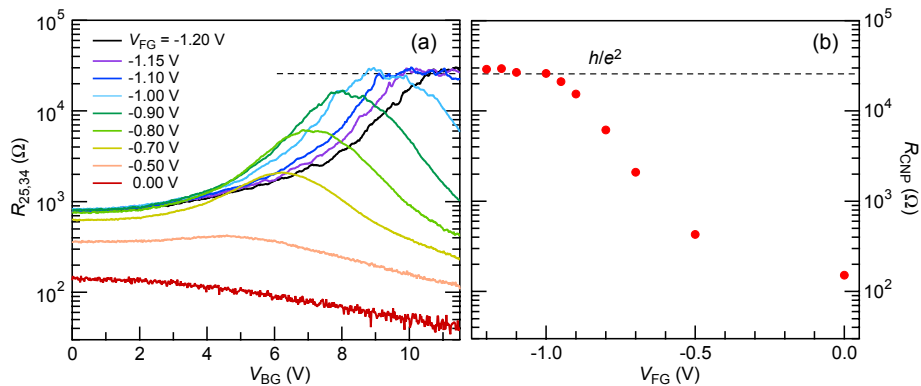}
   \caption{(a) Back-gate voltage dependence of the four-terminal resistance $R_{25,34}$ at different $V_{\textrm{\scriptsize{FG}}}$. (b) Evolution of the corresponding resistance peak amplitude at the CNP as a function of $V_{\textrm{\scriptsize{FG}}}$. In both (a) and (b), the horizontal dashed lines represent $h/e^{2}$.\label{fig3}}
 \end{figure*} 
 
We first describe the evolution of bulk transport with the gate electric field. This is done by using the two-terminal geometry shown in Fig.~\ref{fig2}(a). We pass current between contacts $2$ and $8$ on opposite edges. Since the length of the mesa edges connecting the source and drain is very long ($\sim1$ mm), the measurement primarily probes conduction through the bulk in this configuration, as in a Corbino geometry. Figure~\ref{fig2}(b) shows the back-gate voltage ($V_{\textrm{\scriptsize{BG}}}$) dependence of the two-terminal resistance $R_{28,28}$ at different fixed front-gate voltage $V_{\textrm{\scriptsize{FG}}}$. At $V_{\textrm{\scriptsize{FG}}} = 0$ V, $R_{28,28}$ is small ($\leq2$ k$\Omega$) and only weakly dependent on $V_{\mathrm{BG}}$, indicating a large conduction through the bulk. When a negative $V_{\textrm{\scriptsize{FG}}}$ is applied, $R_{28,28}$ increases and exhibits a peak indicating the charge-neutrality point (CNP) at which the majority carriers cross over from holes on the low-$V_{\textrm{\scriptsize{BG}}}$ side to electrons on the high-$V_{\textrm{\scriptsize{BG}}}$ side. Here, the effect of negative $V_{\textrm{\scriptsize{FG}}}$ is twofold: it lowers the electron density and also reduces the degree of band overlap $E_{g0}$ [Fig.~\ref{fig1}(c)] \cite{suzuki2015}. As $V_{\mathrm{FG}}$ is tuned more negative, the amplitude of the
peak increases and reaches $10$ M$\Omega$ at $V_{\mathrm{FG}}=-1$~V,
demonstrating a good insulation of the bulk. The large change in $R_{28,28}$,
by almost four orders of magnitude, highlights the high gate tunability of the
bulk conduction in our sample. Suppression of residual bulk conduction by
the electric-field effect has previously been reported for a Be-modulation-doped
InAs/GaSb heterostructure and interpreted in terms of the variation of the effective gap with the degree of band overlap \cite{suzuki2015}. The data in Fig.~\ref{fig2}(b) show that a
similar electric-field effect also works for an undoped sample in which
localization of bulk electronic states due to remote-impurity potential is
considered to be weaker.

Now we examine edge transport by passing current between contacts $2$ and $5$
on the same edge and probing voltages at contacts $3$ and $4\ $(separated by
$4$ $\mu$m) between them. Figure~\ref{fig3}(a) shows the $V_{\mathrm{BG}}$ dependence
of the four-terminal resistance $R_{25,34}$, taken at different
$V_{\mathrm{FG}}$. At $V_{\mathrm{FG}}=0$~V, $R_{25,34}$ is small ($\leq 200 $
$\Omega$). With application of negative $V_{\mathrm{FG}}$, it increases and
exhibits a peak at the CNP. This behavior is similar to that of $R_{28,28}$
and hence can be understood to reflect the gate-voltage dependence of the bulk
transport characteristics. However, as $V_{\mathrm{FG}}$ is tuned more negative, the amplitude of the $R_{25,34}$ peak around the CNP saturates for $V_{\mathrm{FG}}\leq-1$~V, indicating the contribution of edge
transport. The $V_{\mathrm{FG}}$ dependence of the resistance peak amplitude
is shown in Fig.~3(b). Note that the resistance saturates close to the quantum of resistance
$h/e^{2}$ (shown by the horizontal dashed line). As we will discuss later,
this is the value expected for quantized edge transport in a single helical edge channel \cite{buttiker1986,roth2009}. 
  
\begin{figure}[!b]
 \centering
  \includegraphics[scale=0.55]{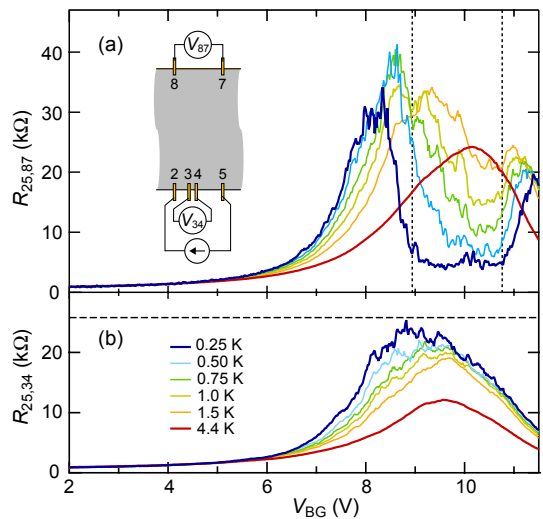}
  \caption{Comparison of four-terminal resistances (a) $R_{25,87}$ and (b) $R_{25,34}$ as a function of $V_{\textrm{\scriptsize{BG}}}$ measured at $V_{\textrm{\scriptsize{FG}}}= -1$ V and at different temperatures. The vertical dashed lines in (a) demarcate the regime where the bulk conduction is suppressed at low temperature. The horizontal dashed line in (b) indicates $h/e^2$. Inset in (a): schematic representation of the measurement configuration. \label{fig4}}
  \end{figure}

 Further evidence that the transport is governed by a single edge
channel is provided by comparing the behavior of the four-terminal resistance
$R_{25,87}$, measured using the voltage probes on the opposite edge, with that
of $R_{25,34}$. Figure~\ref{fig4} compares the $V_{\mathrm{BG}}$ dependence of (a)
$R_{25,87}$ and (b) $R_{25,34}$ measured simultaneously at $V_{\mathrm{FG}%
}=-1$~V. The comparison is made at various temperatures from $0.25$ to $4.4$ K. As the temperature is raised, the peak amplitude of $R_{25,34}$ drops
rapidly above $1.5$ K, indicating significant bulk conduction at elevated
temperatures. This is also confirmed by noting that at $T=4.4$ K, $R_{25,87}$
and $R_{25,34}$ show similar $V_{\mathrm{BG}}$ dependence. This is as expected
for the bulk-dominated transport regime, where the ratio between resistances
measured in different configurations is determined simply by a geometrical
factor. This near proportionality between $R_{25,87}$ and $R_{25,34}$ is seen
to also hold at lower temperatures down to $0.25$~K, but only in those regions not in the vicinity of the CNP. Near the CNP, a dip appears in $R_{25,87}$ at $1.5$~K, which develops into a deep and broad minimum at lower
temperatures. The $R_{25,87}$ minimum saturates at $\sim 5$ k$\Omega$. By comparing this value with the resistance of
the edge segment 8-7 measured in the two-terminal configuration ($\sim 750$
k$\Omega$), the current flowing through the upper edge can be estimated to be
$\sim$ 0.7$\%$ of the injected current in the configuration shown in Fig.~\ref{fig4}. This demonstrates the suppression of conduction along the longer edge in the QSHI phase. This result is consistent with the characteristic edge scattering length of a few microns estimated from the length dependence of the edge resistance \footnote{Two-terminal resistance measured for different edge lengths (26, 180, 260, and 750 k$\Omega$ for 4, 10, 20, and 38 $\mu$m) gives a characteristic edge scattering length ranging from 1.5 to 4 $\mu$m. The 2-mm-long edge is therefore sufficiently resistive to suppress conduction.}. In the following, we focus on measurements at $T=0.25$ K, where the bulk conduction is suppressed\footnote{The temperature dependence of the bulk conduction in Fig.~\ref{fig4}(a) provides a crude estimate of the energy gap $\sim$ 0.25 meV, which is much smaller than the expected hybridization gap of a few meV. We believe that this gap reduction is due to disorder.}.

\begin{figure}[!t]
  \includegraphics[scale=0.6]{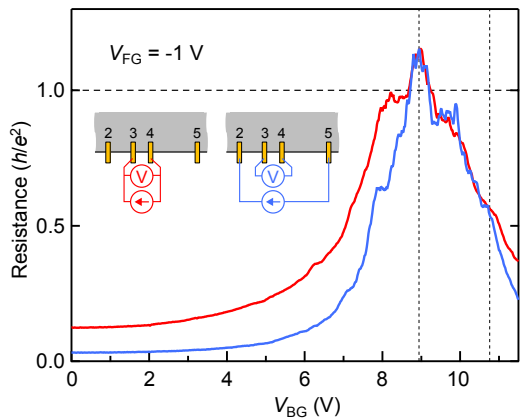}
  \caption{Back-gate voltage dependence of two-terminal resistance $R_{34,34}$ (red) and four-terminal resistance $R_{25,34}$ (blue) measured at $V_{\textrm{\scriptsize{FG}}}= -1$ V. The vertical dashed lines indicate the bulk insulating regime shown in Fig.~\ref{fig4}(a). The inset shows the respective measurement configurations. \label{fig5}}
  \end{figure}

 An advantage of our single-edge configuration is that it allows us to examine the influence of voltage probes on the measured resistance by comparing
two-terminal and four-terminal measurements, as has been done for conventional
1D conductors \cite{bezryadin1998,picciotto2001,makarovski2007}. Figure~\ref{fig5} shows the resistance of segment $3$-$4$ measured in the two-terminal ($R_{34,34}$) and four-terminal ($R_{25,34}$) configurations at $V_{\mathrm{FG}}=-1$~V. (The two-terminal resistance includes the series resistance, which is
estimated to be below 1 k$\Omega$.) Away from the CNP ($V_{\mathrm{BG}}<8$~V), where the
current flows mostly through the bulk, the four-terminal configuration gives a
much lower resistance value, reflecting the more dispersed current
distribution. In contrast, the two configurations give the same resistance value over
a range of $V_{\mathrm{BG}}$ near the CNP, which coincides with the region where the
bulk conduction is suppressed (indicated by the vertical dashed lines). This implies that all the current flowing from contact $2$ to $5$ on the lower edge passes through voltage probes $3$ and $4$. While this is additional evidence for edge-dominated transport, its more important implication is that all the edge modes are absorbed and
thus equilibrated in the voltage probes; that is, contacts $3$ and $4$
behave as ideal reservoirs with perfect absorption for incoming edge modes. 

As already noted, with the Landauer-B\"{u}ttiker model the resistance measured
on a \textit{single} ballistic helical edge channel is expected to be
$h/e^{2}$. This is true for both two-terminal and four-terminal configurations. Importantly, in both configurations the resistance of $h/e^{2}$ arises from dissipation in the Ohmic contacts used as the voltage probes. Inelastic scattering process that occurs \textit{between} the voltage probes would also equilibrate the left-moving and right-moving modes, resulting in an excess resistance. Thus, while the observed saturation of resistance near $h/e^{2}$ (Fig.~\ref{fig3}) suggests nearly ballistic edge transport, several issues require consideration. In particular, the fluctuations of the resistance indicate scattering along the edge. Our single-edge configuration allows us to associate it with scattering centers located near the relevant edge, while unambiguously ruling out processes in other edge segments or bulk-mediated scattering between opposite edges. This is corroborated by the fact that the two-terminal and four-terminal resistances agree with each other, including the fluctuations (Fig.~\ref{fig5}). Interestingly, $R_{25,34}$ does not necessarily agree with $R_{34,34}$ even when the latter takes values very close to $h/e^{2}$ (see the low-$V_{\mathrm{BG}}$ side of the CNP). On the other hand, the matching between $R_{25,34}$ and $R_{34,34}$ continues on the high-$V_{\mathrm{BG}}$ side of the CNP, where they both drop significantly below $h/e^{2}$. A possible reason for this is a potential inhomogeneity or band bending near the sample edge, which may cause an $n$-type region to develop near the edge before the bulk starts to conduct. While further studies are needed to clarify the precise mechanism of the non-ideal behavior, the above results highlight the advantage of performing transport measurements in our single-edge configuration, which allows us to access microscopic electronic properties of 2D TIs from local measurements, complementary to spatially resolved studies performed by scanning gate microscopy \cite{konig2013}.

In summary, we have investigated the transport properties of a band-inverted InAs/GaSb heterostructure using a new geometry in which only a single edge channel is involved in the QSHI phase. By using the electric-field effect, we have suppressed the residual bulk conduction and observed a nearly quantized resistance plateau indicating quasiballistic transport. Our results are promising from the perspective of engineering low-dimensional helical conductors made from 2D TIs. Our approach allows, for instance, for more flexible designs for sophisticated experiments based on QSHI-superconductor hybrid structures.

\begin{acknowledgments}
We sincerely thank Y. Ishikawa and H. Murofushi for their help during the sample preparation.
This work was supported by JSPS KAKENHI Grants No. 26287068 and No. 15H05854.
\end{acknowledgments}

\providecommand{\noopsort}[1]{}\providecommand{\singleletter}[1]{#1}%

\end{document}